\documentclass[aps,prl,twocolumn,superscriptaddress,showpacs]{revtex4}
\usepackage{color}
\usepackage{amsmath}
\usepackage{verbatim}
\usepackage{hyperref}
\usepackage{epsf}
\usepackage{graphicx}
\usepackage{dcolumn}
\usepackage{bm}

\begin{document}
\title{Releasing half-quantum vortices via the coupling of spin polarization, charge- and spin-current}
\author{Hae-Young Kee}
\email{hykee@physics.utoronto.ca}
\affiliation{Department of Physics, University of Toronto, Toronto,
Ontario M5S 1A7 Canada}
\affiliation{Canadian Institute for Advanced Research, Toronto, Ontario  Canada}
\author{Manfred Sigrist}
\email{sigrist@itp.phys.ethz.ch}
\affiliation{ Theoretische Physik, ETH Zurich,
CH-8093 Zurich, Switzerland}

\begin{abstract} 
Motivated by recent experiments observing half-integer flux quanta in mesoscopic loops of superconducting Sr$_2$RuO$_4$, a theory is developed based on
a ``$d$-soliton", a topological defect of spin-triplet superconductors involving a change of the spin configuration. A phenomenological Ginzburg-Landau free energy is given which allows to describe such a soliton yielding a half-integer flux quantum $ (n+1/2) \Phi_0$ inside a loop, assuming a chiral p-wave superconducting phase. The $d$-soliton couples to spin polarization perpendicular to the loop axis, which allows to reduce the energy of the $d$-soliton, fitting well to the experimental findings. The origin of this coupling lies in the combination of spin and charge current within the $d$-soliton generating  a spin polarization analogous to a spin Hall effect. The effect of Fermi liquid corrections is discussed, in particular, in view of the spatial extension of the $d$-soliton and the 
energetics in comparison with a standard integer flux state. 
\end{abstract}


\maketitle

\textit{Introduction -}  The discovery of superfluid phases of $^3$He in 1971 had generated fascinating ideas in modern theoretical physics,
including the simultaneous development of  several broken symmetries and a multitude of topological defects associated with the complex order parameters \cite{vollhardt,leggett-rmp}.
In particular, a possibility of the
half-quantum vortex (HQV) in the so-called A-phase was proposed by Volovik and Mineev in 1976 \cite{volovik76jetp,salomaa87rmp}, 
where the vortices carry a half-integer of the winding number $N= \pm 1/2$. Such type of vortices are of interest also as they can host
quasiparticles of non-Abelian statistics\cite{ivanov01} desirable in the context of quantum computation.
Interestingly, the chiral p-wave spin-triplet equal spin pairing (ESP) state was proposed \cite{sigrist95jpcm} for superconductivity discovered in  Sr$_2$RuO$_4$ \cite{mackenzie03rmp}, which motivated  further search for HQV and studies on unconventional superconductivity in correlated electronic materials. 

The spin-triplet order parameter is generally given as the gap functions,
\begin{equation}
\Delta_{\alpha \beta} ({\bf k}) = {\vec d}({\bf k}) \cdot ({\vec \sigma} i \sigma_y)_{\alpha \beta}.
\end{equation}
Here the d-vector conveys a particular pairing symmetry. For instance, the proposed chiral p-wave
ESP state is represented by
\begin{equation}
{\vec d}({\bf k}) = \Delta_0 {\hat z} \left( k_x \pm i k_y \right),
\end{equation}
where $\sigma_\mu$$ (\mu=x,y,z)$ are Pauli matrices in spin space, and thus $\alpha, \beta$ represent spin $\uparrow$ and $\downarrow$. 
The wave vector $k_\nu$ $(\nu = x,y)$ describes the projection of the unit wave vector ${\hat k}$ along two perpendicular directions such as ${\hat x}$ and ${\hat y}$
in two dimensional space. This order parameter represents the Cooper pair with the zero-spin projection on ${\hat z}$ (d-vector direction), and thus
up-up (down-down) Cooper pair orientation is perpendicular to the z-axis.
The orbital angular momentum of the Cooper pairs is given by ${\hat l} = {\hat x} \times {\hat y} \parallel {\hat z}$ pinned due to the spin-orbit coupling. 
Since it is characterized by the p-wave internal structure  and spin triplet condensates represented by ${\hat d}$-vector,
the HQV then can be viewed as the sign change in  the angular momentum of Cooper pair ${\hat l}$-vector (phase of angular momentum), and this change in sign is 
compensated by a concurrent rotation of ${\hat d}$-vector into $-{\hat d}$ \cite{salomaa87rmp,kee00prb} .

While the evidence of its chirality $(k_x \pm i k_y)$ in Sr$_2$RuO$_4$ is still under the debate \cite{kallin12rpp}, a recent report on the observation of half-height magnetization steps \cite{jang11science} is
compatible with the spin-triplet pairing supporting the existence HQV.
Theoretically, the free energy analysis of HQV studied in $^3$ He-A phase\cite{vollhardt} was applied to Sr$_2$RuO$_4$ in Ref.\cite{kee00prb}.
The energetic balance between the HQV and single-vortex
was investigated, and a pair of half quantum vortices (HQVs) with an optimized separation was found to be energetically stable in a certain range of the magnetic field (see Fig.\ref{loop} a) \cite{kee00prb},
Later, a micro-size sample for the stability of the HQV was suggested \cite{chung07prl}, and it was pointed out that the HQVs accompany a finite spin polarization that could be an experimental
indicator of the HQV \cite{kee07epl,vakaryuk09prl}. 
Motivated by these studies,  Jang {\it et al} designed mesoscopic samples of Sr$_2$RuO$_4$, and performed
cantilever magnetometry measurements on micrometer annular shaped samples \cite{jang11science}. Half-height magnetization steps were found in the presence of
an external in-plane magnetic field, supporting the HQV in Sr$_2$RuO$_4$.
This surprising experimental finding calls for a theoretical explanation that incorporates the effect of in-plane magnetic field and its relation to spin polarization and the HQV.

In this paper, we investigate the free energy of the HQV including the spin polarization effect.  The in-plane magnetic field permits the spin polarization which in turn
allows the coupling between spin and charge currents of Cooper pairs. This coupling indeed lowers the free energy of the HQV state compared with the single vortex phase, revealing the HQV in the ring shaped Sr$_2$RuO$_4$ superconductor.

\textit{Free energy of the HQV -} 
 Based on the symmetry analysis, we propose a simple free energy, which is sufficient to capture the relevant features,
\begin{eqnarray}
F[\vec{d}] &=& \int d^3r \; \left[ K \sum_{j=x,y,z} \left| D_j \vec{d} \, \right|^2 +  \lambda \Delta_0^2 (1 - |\hat{\Delta} \cdot \hat{z}|^2) \right. \nonumber\\
&& \left. + \gamma  \left\{ i (\vec{D} \times \vec{d} \, )^* \cdot (\vec{D} \times (\vec{S} \times \vec{d}) \,) + c.c. \right\} \right],
\label{fren-1}
\end{eqnarray}
where $ \vec{D} = i \vec{\nabla} - 2 \pi \vec{A} / \Phi_0$ .
Here  the flux quantum $\Phi_0 = hc/(2 e)$ and $\vec{A}$ is the vector potential.
$ K, \lambda$, and $\gamma $ are real coefficients.
The first term describes the stiffness of order parameter,  and the second term represents the energy cost of tilting the $d$-vector away from the $z$-axis, pinned by
the spin-orbit coupling \cite{kee00prb}. When Fermi liquid corrections are taken into account, the stiffness $K$ differs for the spatial variation of
charge and spin of Cooper pair, which we will discuss later.
The last term is a coupling between charge current, spin current, and spin polarization ${\vec S}$. 
Note that a similar coupling has been discussed in the context of Josephson junction between two spin-triplet superconductors in Ref.\cite{asano06,brydon10}. 
In the following we assume that the spin polarisation $ \vec{S} $ has no spatial dependence. 

To estimate the free energy of HQV and to understand the effects of the last term in the free energy, 
we first relate the coefficients $\gamma$ and $K$ in the weak-coupling limit. 
For this purpose we use a $d$-vector in the $x$-$y$ plane to express the superfluid density in equal-spin pairing state with spin along the $z$-axis.
We consider a gap function with a spatial variation along the $x$-axis,
\begin{eqnarray}
\vec{d}(x) & = & \Delta_0 e^{i q x} \left\{ \hat{x} \cos(q' x) + \hat{y} \sin(q'x) \right\} \nonumber\\ & = & \vec{d}_{0 \uparrow} e^{i(q+q')x} + \vec{d}_{0 \downarrow} e^{i(q-q')x}
\end{eqnarray}
with $ \vec{d}_{0 s} = \Delta_0 (\hat{x} - is \hat{y} ) $ where $s= \pm 1$ for $\uparrow$- and $\downarrow$-spin.
 Inserting this into Eq.(\ref{fren-1}), neglecting spin-orbit coupling ($\lambda =0 $), we find
\begin{eqnarray}
F &=& \int d^3r \Delta_0^2 \left[ K \left\{ (q+ a_x)^2 + {q'}^2 \right\} + \gamma S_z q'(q+a_x) \right]  \nonumber \\
&=& \int d^3r \tilde{K} \left[  \rho_{\uparrow} (q+a_x+q')^2 +   \rho_{\downarrow} (q+a_x-q')^2  \right]
\end{eqnarray}
with $a_x = 2 \pi A_x / \Phi_0 $, where the second line is expressed as separate contributions of up- and down-spin Cooper pairs with $ \tilde{K} $ as a
phenomenological coefficient and $ \rho_s $ the density of electrons for spin $s$.  It is straightforward to identify the coefficient within the two representations using $ S_z = \frac{\rho_{\uparrow} - \rho_{\downarrow}}{2}\hbar $, 
\begin{equation}
K \Delta_0^2  =  \tilde{K} \rho \quad \mbox{and} \quad \gamma = \frac{2 K}{S_0}
\end{equation}
with $ \rho = \rho_{\uparrow} + \rho_{\downarrow} $ and $ S_0 = \hbar \frac{\rho}{2} $.  

\begin{figure}
\includegraphics[width=1.0\columnwidth,angle=0]{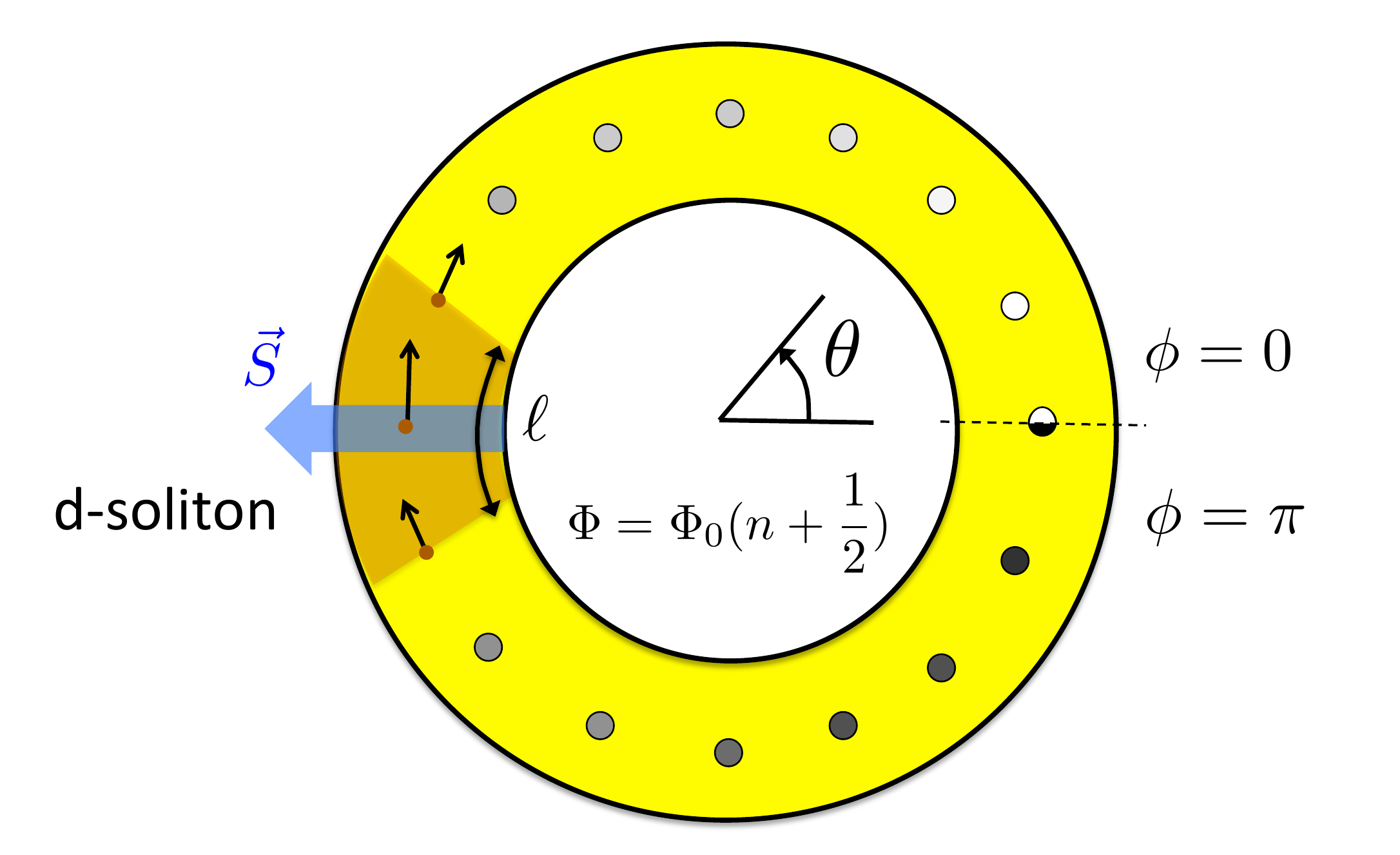}
\caption{Schematic visualisation of a d-soliton. In the region of the d-soliton the d-vector turns from ${\hat z}$ smoothly to ($-{\hat z}$). The circles indicate the d-vector point along the $z$-axis (parallel / antiparallel). The gray scale of the circles indicate the phase $ \phi $ continuously changing from $ \phi =0 $ to $ \phi = \pi $.  $\Phi$ denotes the half-integer flux quantum captured in the center of the ring (here $ n = 0 $). The blue arrow 
marks the spin polarization in the d-soliton due to the superposition of charge and spin current.}
\label{flux-loop}
\end{figure}

We now include spin-orbit coupling ($ \lambda > 0 $) in Eq.(\ref{fren-1}) and consider the ring geometry of Fig.\ref{flux-loop}. The stable phase has $ \vec{d} \parallel \hat{z} $. In order to examine order parameter windings, we follow Fig. 1 and limit ourselves to the following ansatz,
\begin{equation}
\vec{d} (\theta)  
= \Delta_0 e^{i \phi_\theta} \{\hat{z} \cos{(\alpha_\theta)} + \hat{\theta} \sin{(\alpha_\theta}) \},
\end{equation}
using cylindrical coordinates $({\hat r}, {\hat \theta}, {\hat z})$. There are two types order parameter windings in this configuration. The standard phase winding is characterised by $ \alpha_{\theta} = 0 $ while $ \phi_{\theta} $ winds by an integer multiple of $ 2 \pi $ (e.g. $\phi_{\theta} = \theta $). The unusual type of winding
is depicted in Fig. 1 where $ \alpha_{\theta} $ varies such that around $ \theta = 0$
the $d$-vector changes smoothly from $ +\hat{z} $ to $ -\hat{z} $ - this is a {\it $d$-soliton} - accompanied by a winding of $ \phi_{\theta} $ by $ (2n+1)\pi $. The former gives rise to a magnetic flux 
of an integer multiple of $ \Phi_0 $, while we see a half-integer flux quantum in the later case. 

For a concrete discussion we keep the order parameter magnitude $ \Delta_0 $ constant and vary only $ \phi_{\theta} $ and $ \alpha_{\theta} $. For the vector potential we use ${\bf A} = A_{\theta}(r,z) \hat{\theta} = \frac{ B_z r}{2}  \hat{\theta}$ such that $\nabla \times {\bf A} = B_r {\hat r} $ and $\nabla \cdot {\bf A} =0$. With this ansatz only the radial part of the spin polarisation appears in Eq.(\ref{fren-1}), $ \vec{S} = S_r {\hat r}$.
The corresponding the free energy functional then reads,
\begin{eqnarray}
F[\alpha_\theta, \phi_\theta] &=& \eta  \int d\theta  \; \left[  K_{\alpha} (\partial_{\theta} \alpha_\theta)^2 + K_{\phi}  (\partial_{\theta} \phi_\theta + a'_{\theta})^2  \right. \nonumber\\
&& \hskip -1.5cm \left.  + \lambda R^2 (1 - \cos^2 \alpha_\theta ) + 4 K_{\alpha \phi} s_r  \partial_{\theta} \alpha _\theta (\partial_{\theta} \phi_\theta + a'_{\theta}) \right],
\label{energy}
\end{eqnarray}
where $\eta = \Delta_0^2  d / R $ and $ s_r = S_r / S_0 $.  We assume homogeneity in radial direction and the width $ d $ of the annulus is much smaller than the radius $ R$.
Note that $a'_\theta \equiv a_\theta  R = 2 \pi A_\theta R/\Phi_0 =  \Phi/\Phi_0$ where $\Phi = B_z \pi R^2$. We introduce individual coefficients for the different gradient terms which allow to include Fermi liquid corrections:
\begin{eqnarray}
K_{\alpha}/K  &=&  1 + F_1^a/3  , \;  K_{\phi}/K =  1 + F_1^s/3 , \nonumber \\ 
K_{\alpha \phi}/K  &=&  1  + F_1^s/3 + F_1^a/ 3
\label{fermi-liquid}
\end{eqnarray}
with $ F_1^s $ and $ F_1^a$ as the Landau parameters renormalizing the charge and spin current, respectively \cite{leggett-rmp}. Note that
usually $ F_1^s > 0 $ and $ F_1^a < 0 $, e.g. in $^3$He \cite{leggett65}. The standard weak coupling approach offers $ K_{\alpha} = K_{\phi} = K_{\alpha \phi} \equiv K $. 

\begin{figure}
\includegraphics[width=1.0\columnwidth,angle=0]{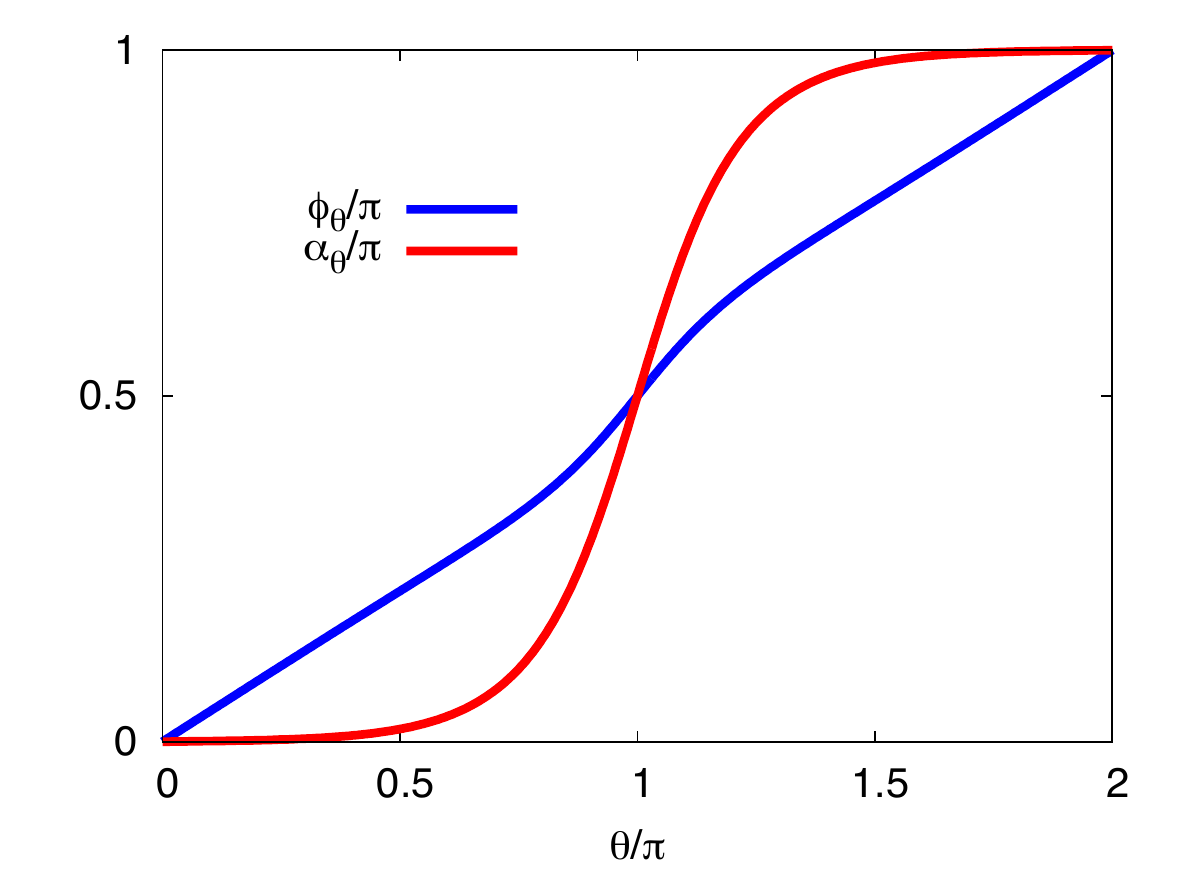}
\caption{The behavior of $\alpha_\theta$ and $\phi_\theta$ are shown as a function of $\theta$ obtained by numerical variation of Eq.(\ref{energy}).
For illustrative purpose we set $\lambda R^2 =5 K_{\alpha} $ with $ K_{\alpha} = K_{\phi} = K_{\alpha \phi} $ (weak-coupling), and $s_r \equiv s_x \cos{\theta}$ with $s_x = 0.125$ fixing the orientation of the spin magnetization to the $x$-axis. The the chosen boundary conditions the $d$-soliton, visible as a smooth step in $ \alpha_{\theta} $ is centred at $ \theta = \pi $. Through the coupling to the spin polarization also the phase $ \phi $ squires a weak anomalous in the region of the $d$-soliton.}
\label{fig-phase}
\end{figure}

\textit{Towards qualitative understanding -} 
The variation of $ F$ in Eq.(\ref{energy}) with respect to $ \alpha_{\theta} $ and $\phi_{\theta} $ leads to
\begin{equation} \begin{array}{l} 
\displaystyle  2 K_{\alpha} \partial_{\theta}^2 \alpha_{\theta}  - \lambda \sin 2 \alpha_{\theta}  +
4 K_{\alpha \phi}s_r \partial_{\theta}^2 \phi_{\theta}  = 0, \\ \\
\displaystyle K_{\phi} \partial_{\theta}^2 \phi_{\theta}  + 2 K_{\alpha \phi} s_r  \partial_{\theta}^2 \alpha_{\theta}  = 0.
\end{array}
\label{var}
\end{equation}
These coupled equations can be solved for a $d$-soliton solution by
\begin{eqnarray}
\alpha_{\theta} &=&  2 \arctan \left[ e^{\theta / \ell} \right],  \\
\phi_{\theta} &=& - \frac{4 K_{\alpha \phi} s_r}{K_{\phi}} \arctan \left[ e^{\theta / \ell} \right]  + c' \theta,
\label{sol-alpha-phi}
\end{eqnarray}
where the soliton is centred at $ \theta = 0 $ and the characteristic length 
\begin{equation}
\ell = \sqrt{ \frac{ K_{\alpha}}{\lambda R^2} \left( 1 - \frac{4 K_{\alpha \phi}^2 s_r^2}{K_{\alpha} K_{\phi} } \right)}
\label{ell}
\end{equation}
is small enough as to guarantee the validity of the periodic boundary conditions sufficiently well ($ \ell \ll 2 \pi  $). 
For a single-valued order parameter it is important to imposes $\alpha_\theta + \phi_\theta = 2 \pi n $ where $n$ is an integer. 
Then the coefficient $c'$ has to satisfy the constraint that $ \phi_{\theta} $ exhibits a half-integer winding, 
\begin{equation}
c' = n+\frac{1}{2} + \frac{K_{\alpha \phi} s_r}{K_{\phi} } .
\end{equation}
Note that for a narrow $d$-soliton the formally radial spin polarization is nearly equivalent to the situation of a uni-directional polarization, since the
coupling is only finite when $ \partial_\theta \alpha_{\theta} \neq 0 $. 

We may give a rough quantitative estimate of $ \ell $ based on a weak coupling approach.  With 
\begin{equation}
K \sim N(\epsilon_F) \xi_0^2 \quad \mbox{and } \quad  \lambda \sim N(\epsilon_F) \frac{\Delta T_c}{T_c},
\end{equation}
where $ N(\epsilon_F) $ is the density of states at the Fermi energy, $ \xi_0 $ is the coherence length at $ T=0 $ and $ \Delta T_c / T_c $ is the relative
change of $ T_c $ when the spin degeneracy of the spin-triplet pairing states is lifted by spin-orbit coupling ($ \Delta T_c / T_c \sim 5 \% $) \cite{ng00,yanase03}. 
Then we find with Eq. (\ref{ell}) 
\begin{equation}
\ell \sim \frac{\xi_0}{R} \sqrt{\frac{T_c}{\Delta T_c}  (1 - 4 s_r^2) },
\end{equation}
where $ \ell \lesssim 1$ which is reduced through the spin polarization.

With the shape of the $d$-soliton in Eq.(\ref{sol-alpha-phi}), we now calculate the energy of the HQV state,
\begin{eqnarray}
F_{HVQ} = 2 \eta \lambda R^2 \ell + 2 \pi \eta K_{\phi} \left\{ \Phi_{\frac{1}{2}}^2 - \frac{2K_{\alpha \phi} s_r}{K_{\phi} \ell}
\Phi_{\frac{1}{2}} \right\}
\label{en-HQV}
\end{eqnarray}
with $ \Phi_{\frac{1}{2}} = (n +1/2) - \Phi/\Phi_0 $. The first term constitutes the energy of $d$-soliton while the second includes the effects of
the currents. 
For the integer quantum vortex (IQV) we use $ \alpha_{\theta} = 0 $ and $ \phi_{\theta} = n \theta $ which 
yields the free energy,
\begin{equation}
F_{IQV} =2 \pi  \eta K_{\phi} \left(n - \frac{\Phi}{\Phi_0} \right)^2 .
\label{en-IQV}
\end{equation}
Now we compare the free energies of two states as a function of the flux $ \Phi $. The flux minimizing $ F_{HQV} $ is given by
\begin{equation}
\frac{\Phi_{\rm min}}{\Phi_0} = n+\frac{1}{2} - \frac{K_{\alpha \phi} s_r}{K_{\phi} \ell}  
\label{Phi-min}
\end{equation}
which deviates from half-integer flux quantisation, if there is a non-vanishing spin polarization.  This result corroborates that the spin polarization induces an additional supercurrent, which appears in 
Fig. \ref{fig-phase} as a change of slope in $ \phi_{\theta} $ around the $d$-soliton resulting in a modified
coefficient $ c' $ for $ \phi_{\theta} $. 

{\it Energy balance between HQV and IQV} - We now examine whether it is possible to stabilize the HQV relative to the IQV phase for some flux $ \Phi $. A straightforward analysis shows
that for small deviations of $ \Phi_{\rm min} $ from half-integer flux quanta ($2 K_{\alpha \phi} s_r < K_{\phi} \ell $), the two states, IQV and HQV, interchange their role as stable phase at $ \Phi = (n+1/2) \Phi_0$. The condition for the HQV phase being lower in energy is given by
\begin{equation}
\frac{\pi^2}{16} \frac{K_{\phi}^2}{K_{\alpha} \lambda R^2} + \frac{4 K_{\alpha \phi}^2 s_r^2}{K_{\alpha} K_{\phi}} > 1 \;.
\end{equation}
The first term on the left-hand side originates from the comparison of the standard fluxoid energy with the $d$-soliton energy which has to be paid for a HQV phase. Note that the condition for a HQV is improved through the Fermi liquid corrections introduced in Eq.(\ref{fermi-liquid}). 
Moreover, it is obvious through the second term on the left-hand side that in-plane spin polarization boosts the HQV. 

The mechanism for this second term can be illustrated by considering the current densities. The charge current density around the cylinder is obtained by variation of $F$ in 
Eq. (\ref{energy}) with respect of the vector potential, $ A_{\theta} $,
\begin{equation}
J_{\theta} = \frac{4e\Delta_0^2}{\hbar} \left\{ K_{\phi} (\partial_{\theta} \phi_{\theta} + a_{\theta}' ) + 2 K_{\alpha \phi} s_r \partial_{\theta} \alpha_{\theta} \right\} \; .
\end{equation}
To derive the spin current we introduce an auxiliary twist on $ \vec{d} $, replacing $ \alpha_{\theta} $ by $ \alpha_{\theta} + a_s \theta $. The variation with respect to $ a_s $ (setting eventually $ a_s$ to zero) yields,
\begin{equation}
J_{\theta}^r =  2 \Delta_0^2 \left\{ K_{\alpha} \partial_{\theta} \alpha_{\theta} + 2 K_{\alpha \phi} s_r (\partial_{\theta} \phi_{\theta} + a_{\theta}' ) \right\}
\end{equation}
for the spin component parallel to $ \hat{r} $. Both currents include contributions coupled to spin polarization, in the sense of a spin Hall effect. 
This coupling indeed provides a reduction of energy by spin polarization. We may view the $d$-soliton and the phase $ \phi $ gradient as a source driving an in-plane spin polarization which can be enhanced through an external in-plane magnetic field. 

With the free energy  Eq.(\ref{fren-1}), we can also discuss another limit in the shape of the $d$-soliton, 
\begin{equation}
\vec{d} (\theta)  
= \Delta_0 e^{i \phi_\theta} \{\hat{z} \cos{(\alpha_\theta)} + \hat{r} \sin{(\alpha_\theta}) \} \;.
\end{equation}
This transverse twist structure of the soliton yields essentially the same equations of $ \alpha $ and $ \phi $ and energy estimates. The relevant spin polarisation is now directed along $ \hat{\theta} $ ($\vec{S} = S_{\theta} \hat{\theta} $), and as long as the extension of the soliton is small compared to the circumference of the cylinder, the spin polarization is restricted to a small range of $ \theta $ such that it is practically uni-directional. In this way any intermediate form of $d$-solitons can be discussed, which leads, however, generally to more complex equations. This is important in view of the possible pinning of $d$-solitons at constrictions 
(weak regions) on the cylinder to reduce the soliton energy. This would facilitate the stabilisation of a HQV state. 

\begin{figure}
\includegraphics[width=1.0\columnwidth,angle=0]{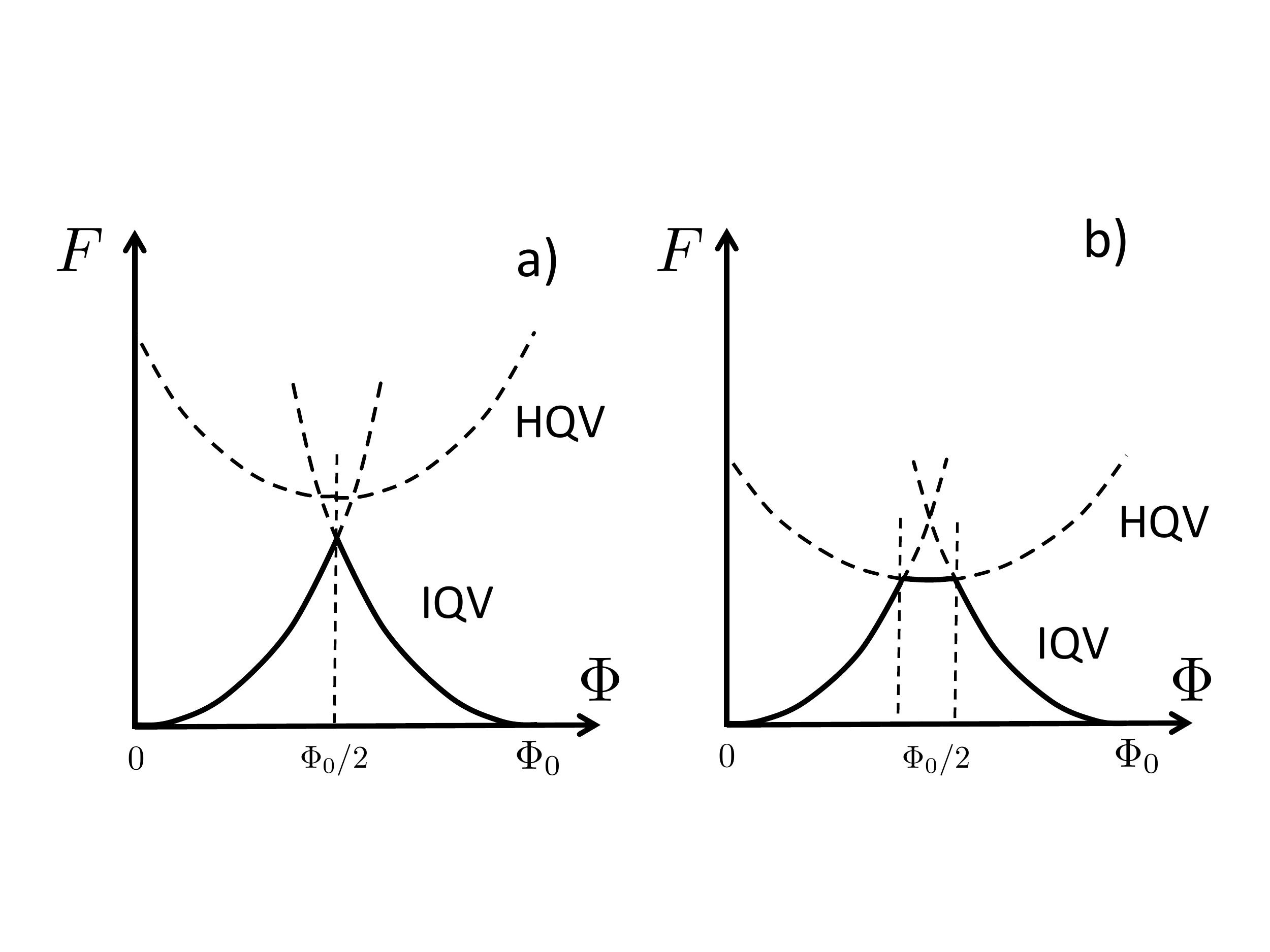}
\caption{Schematic flux dependence of the energies for HVQ and IVQ phases between $ \Phi = 0 $ and $ \Phi_0 $ based Eqs.(\ref{en-HQV}) and (\ref{en-IQV}): a) The minimal energy of the HQV phase at $ \Phi=\Phi_0/2 $ lies higher than that of the IQV phase for all fluxes. b) the energy of the HQV phase is lowered such that a window of stability appears around $\Phi=\Phi_0/2 $. Note that this figure can be periodically continued.}
\label{Phi}
\end{figure}

\begin{figure}
\includegraphics[width=1.0\columnwidth,angle=0]{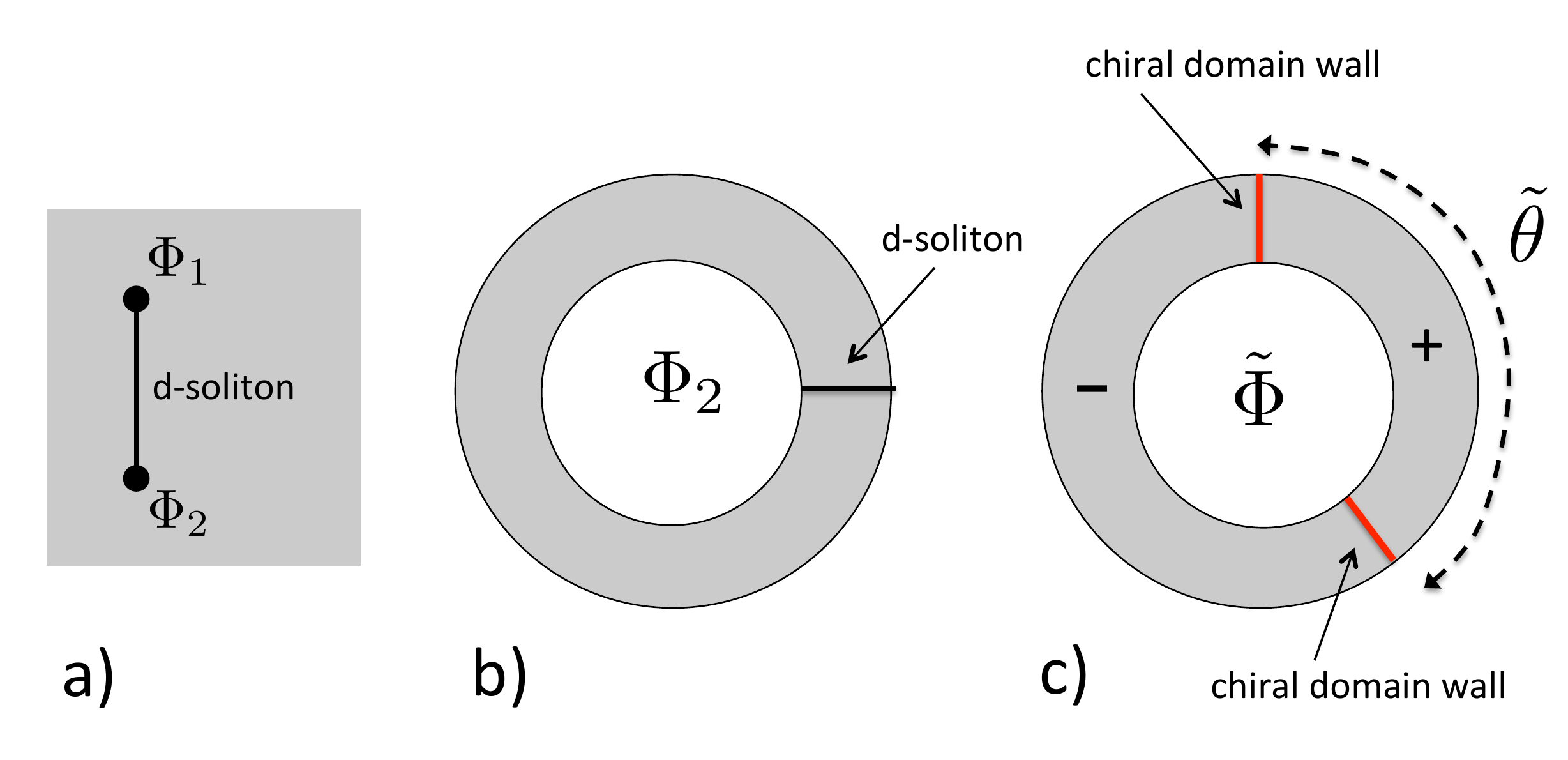}
\caption{Configurations of non-integer fluxes: a) top view on a planar $d$-soliton in a bulk superconductor connecting two half-integer flux lines, $ \Phi_1 $ and $ \Phi_2 $ with $ \Phi_1 + \Phi_2 = n \Phi_0 $;  b) HQV phase of cylinder corresponds to $d$-soliton connecting a half-integer flux $ \Phi_2 $ in the interior with the outside; c) alternative configuration with two domain walls separating two chiral domains and a flux depending of the position of the domain walls; $ \tilde{\Phi} $ depends on details of the configuration\cite{bouhon10,ferguson11}.}
\label{loop}
\end{figure}

Equipped with this phenomenology, the d-solition constitutes a possible explanation for the observation of half-quantised fluxes in a cylinder of Sr$_2$RuO$_4$. The HQV window is opened and widened by an in-plane magnetic Zeeman field in agreement with our finding that a $d$-soliton can be stabilised by an in-plane spin polarization. This in turn provides a narrowing of the soliton shape. 
Thus, without spin polarization, the HQV state may be too high in energy to appear as a stable state at any magnetic flux $ \Phi $ through the cylinder, realizing standard flux quantisation as shown in Fig. \ref{Phi}(a).
However, lowering the soliton energy by spin polarization can yield a flux range where HQV state is stabilized as shown in Fig. \ref{Phi}(b). 

Introducing a $d$-soliton into the cylinder is equivalent to a phase slip between standard flux quanta in a cylinder, i.e. the passage of a vortex through the cylinder wall. A $d$-soliton in the bulk of a superconductor can, in principle appear like a planar defect terminated at both end by magnetic flux lines of half-integer flux quanta as depicted in Fig.\ref{loop}(a). Thus, a half-flux quantum passing the cylinder wall drags a $d$-soliton behind presented in Fig.\ref{loop}(b).  Then the following half-flux removes the soliton, and adds the 
total flux to an integer multiple of $ \Phi_0 $. As noted above, a constriction in the cylinder wall may play an important role too. A vortex enters usually 
through weakest region of the loop, e.g. a constriction. This is also beneficial for the half-flux vortex and in addition the energy expense for the
$d$-soliton would be reduced. Thus, it may even be possible to stabilize the HQV state without in-plane magnetic field, if there is a constriction.  

An alternative way to obtain a fractional flux quantum in a cylinder geometry is the possibility of domain walls, since the chiral $p$-wave state allows for two degenerate domains. As shown in Fig. \ref{loop}(c) this requires two domain walls separating the phase ``+'' $(k_x + i k_y)$ from ``-'' $(k_x - i k_y)$. The flux introduced depends on the angle between the two domain walls \cite{bouhon10,ferguson11}. Note that these fluxes have no fixed value but depend on details. Moreover, it is more involved to generate this state, as it requires to nucleate of domains and to move two domain walls apart. Thus the $d$-soliton way of introducing HQV described above is more natural to obtain a sequence of IQV and HQV.

 \textit{Summary and Discussion} - We presented here a scenario qualitatively consistent with the phenomenology of the observed half-quantum flux vortices in the recent experiments on tiny cylinders of Sr$_2$RuO$_4$. In particular, our analysis explains the role of the in-plane magnetic field in facilitating the 
appearance of the HQV phase, although it is difficult to give a reliable quantitative assessment of our theory at the present stage, as Fermi liquid corrections are not known and the treatment of sample geometrical aspects have also to be considered. An attractive aspect of the combination of the charge and spin currents is the induced spin polarization even in the absence of in-plane magnetic fields. Thus the $d$-soliton in the loop structure is intrinsically ferromagnetic, a feature which can be enhanced for systems close to a magnetic instability \cite{brydon10}.  
At this point we cannot give an analysis of the fact that the recent data by Cai {\it et al} do not display half-flux oscillation in their magneto-resistance measurements of similar loops, apart from the absence of in-plane magnetic fields in this experiment \cite{cai12,vakaryuk12}. It may be interesting in this context to experiment with loops that possess well-defined constrictions where a $d$-soliton could be more stable and pinned.  


 %
\textit{Acknowledgement -}
We are grateful for helpful discussion to R. Budakian, A. Bouhon, P. Brydon, D. van Harlingen, D.A. Ivanov, A.J. Leggett, Y. Liu and V. Vakaryuk.
HYK was supported by NSERC of Canada and MS by the Swiss Nationalfonds and the NCCR MaNEP.


\begin{thebibliography}{99}
\bibitem{vollhardt} D. Vollhardt and P. Wolfle, The superfluid phases of Helium 3 (Taylor \& Francis, New York 1990).
\bibitem{leggett-rmp}  A. Leggett, Rev. Mod. Phys. {\bf 47}, 331 (1975).
\bibitem{volovik76jetp} G. E. Volovik and V. P. Mineev, JETP Lett. {\bf 24}, 561 (1976).
\bibitem{salomaa87rmp} M. M. Salomaa and G. E. Volovik, Rev. Mod. Phys. {\bf 59}, 533 (1987).
\bibitem{ivanov01} D.A. Ivanov, Phys. Rev. Lett. {\bf  86}, 268 (2001). 
\bibitem{sigrist95jpcm} T. M. Rice and M. Sigrist, J. Phys. Cond. Matter, {\bf 7}, L643 (1995).
\bibitem{mackenzie03rmp} A. P. Mackenzie and Y. Maeno, Rev. Mod. Phys. {\bf 75}, 657 (2003).
\bibitem{kee00prb} H.-Y. Kee, Y. B. Kim, and K. Maki, Phys. Rev. B {\bf 62},  R9275 (2000).
\bibitem{kallin12rpp} C. Kallin, Rep.  Prog. Phys. {\bf 75}, 042501 (2012) and references therein.
\bibitem{jang11science} J. Jang, D. G. Ferguson, V. Vakaryuk, R. Budakian, S. B. Chung, P. M. Goldbard, and Y. Maeno,
Science {\bf 331} 186 (2011).
\bibitem{chung07prl} S. B. Chung, H. Bluhm, E.-A. Kim, Phys. Rev. Lett. {\bf 99}, 197002 (2007).
\bibitem{kee07epl} H.-Y. Kee and K. Maki, Europhys. Lett. {\bf 80}, 46003 (2007).
\bibitem{vakaryuk09prl} V. Vakaryuk and A. J. Leggett, Phys. Rev. Lett. 103, 057003 (2009).
\bibitem{asano06} Y. Asano, Phys. Rev. B {\bf 74}, 220501(R) (2006)
\bibitem{brydon10} P.M.R. Brydon, C. Iniotakis, D. Manske and M. Sigrist, Phys. Rev. Lett. {\bf 104}, 197001 (2010).
\bibitem{leggett65} A.J. Leggett, Phys. Rev. {\bf 140}, A1869 (1965). 
\bibitem{ng00} K.-K. Ng and M. Sigrist, Europhys. Letts. {\bf 49}, 473 (2000).
\bibitem{yanase03} Y. Yanase and M. Ogata, J. Phys. Soc. Jpn. {\bf 72}, 673 (2003).
\bibitem{bouhon10} A. Bouhon and M. Sigrist, New J. Phys. {\bf 12}, 043031 (2010)
\bibitem{ferguson11} D.G. Ferguson and P.M. Goldbart, Phys. Rev. B {\bf 84}, 014523 (2011). 
\bibitem{cai12} X. Cai, Y. A. Ying, N. E. Staley, Y. Xin, D. Fobes, T. Liu, Z. Q. Mao, and Y. Liu, ArXiv e-prints (2012), cond-mat-1202.3146.
\bibitem{vakaryuk12} V. Vakaryuk, K. Roberts, D. G. Ferguson, J. Jang, R. Budakian and S. B. Chung, ArXiv e-prints (2012), cond-mat-1203.5771. 

\end{thebibliography}
\end{document}